\documentclass[conference]{IEEEtran}
\IEEEoverridecommandlockouts
\usepackage{amsmath,amssymb,amsfonts,MnSymbol}
\usepackage{algorithmic}
\usepackage{graphicx}
\usepackage{textcomp}
\usepackage{xcolor}
\usepackage{comment}
\usepackage{subcaption}
\usepackage[style=ieee, citestyle=numeric-comp, backend=biber]{biblatex}
\usepackage{relsize}
\usepackage{float}
\usepackage{nopageno}
\usepackage[backend=biber]{biblatex}
\usepackage[%
    colorlinks=true,
    pdfborder={0 0 0},
    linkcolor=blue
]{hyperref}
\def\BibTeX{{\rm B\kern-.05em{\sc i\kern-.025em b}\kern-.08em
    T\kern-.1667em\lower.7ex\hbox{E}\kern-.125emX}}

\DeclareRobustCommand*{\IEEEauthorrefmark}[1]{%
  \raisebox{0pt}[0pt][0pt]{\textsuperscript{\footnotesize\ensuremath{#1}}}}

\begin{document}

\title{Scheduling in Wireless Networks using Whittle Index Theory}

\author{
\IEEEauthorblockN{ Karthik GVB\IEEEauthorrefmark{1}, Vivek S.\ Borkar\IEEEauthorrefmark{2}, Gaurav S.\ Kasbekar\IEEEauthorrefmark{3}}
\\
\IEEEauthorblockA{\IEEEauthorrefmark{1,2,3}Department of Electrical Engineering, Indian Institute of Technology (IIT) Bombay, India}\\
E-mail addresses: \IEEEauthorrefmark{1}karthikgvb@gmail.com,
\IEEEauthorrefmark{2}borkar@ee.iitb.ac.in, \IEEEauthorrefmark{3}gskasbekar@ee.iitb.ac.in.
}

\maketitle
\thispagestyle{empty}
\begin{abstract}
We consider the problem of scheduling packet transmissions in a wireless network of users while minimizing the energy consumed and the transmission delay. A challenge is that transmissions of users that are close to each other mutually interfere, while users that are far apart can  transmit simultaneously without much interference. Each user has a queue of packets that are transmitted on a single channel and mutually non interfering users reuse the spectrum. Using the theory of Whittle index for cost minimizing restless bandits, we design four index-based policies and compare their performance with that of the well-known policies: Slotted ALOHA, maximum weight scheduling, quadratic Lyapunov drift, Cella and Cesa Bianchi algorithm, and two Whittle index based policies from a recently published paper. We make the code used to perform our simulations publicly available, so that it can be used for future work by the research community at large.
\end{abstract}


\section{Introduction}
Recall that multiple users in a wireless network can transmit data simultaneously on the same channel if they are far apart since there is little mutual interference. This is known as spatial reuse of spectrum~\cite{1} and is useful for increasing the capacity of wireless networks~\cite{1}. In the simplest case, these mutually far apart users constitute an independent set~\cite{anderson2001graph} of users in the conflict graph~\cite{anderson2001graph} corresponding to the network~\cite{anderson2001graph}. By scheduling, we mean that in every slot of transmission, a mutually independent/ non-interfering set of users is selected to transmit. A challenge is to accomplish this with low average energy consumption and data transmission delay.

In this paper, we consider a set of transmitter-receiver pairs of users in a region and novel scheduling policies by modeling the wireless network of users using the \emph{restless bandit framework}~\cite{23}. The queue of each user evolves with time and the constraint is to choose a mutually non-interfering set of users. In a restless bandit formulation, the \emph{Whittle index theory}~\cite{23} provides a way to select which users to schedule ($\approx$ arms to activate in the restless bandit formulation) for minimizing the time-averaged cost. Based on the Whittle index theory, we design four policies that show better performance in terms of average cost and average throughtput, than several well-known policies in most of the scenarios considered.

We represent the wireless network by an undirected graph~\cite{anderson2001graph} (fig-\ref{Net}), the nodes represent the users, and there is an edge between two nodes if the corresponding users interfere when transmitting data. We model two costs-- ``energy cost'' and ``holding cost'': energy cost refers to the energy consumed for transmitting packets, which increases with the number of packets transmitted. Holding cost is proportional to the delay incurred and hence is proportional to the queue length. The cost incurred in a slot at a user is the sum of the energy cost and holding cost. The problem is to reduce the time-averaged total cost incurred by all users in the network. We use Whittle index theory~\cite{23} to solve this problem. As in Whittle's theory, we relax the hard constraint of an independent set of users transmitting in each time slot to a time-averaged constraint and formulate a corresponding unconstrained problem using Lagrange multipliers, as the original constraint makes the problem provably hard~\cite{36}. As in Whittle's theory, we decouple the unconstrained problem into individual problems for each user and define suitable Whittle-like indices. A distributed algorithm is also proposed for deciding who should transmit based on the  indices for all the users.

We now provide a review of related prior literature. Scheduling in wireless networks with the objectives of minimizing the energy consumption and/ or delay has been extensively studied in prior work. A survey of schemes for delay-aware resource control in a multi-hop wireless network is provided in~\cite{15}. A scheduling scheme for minimizing the energy-expenditure in a time-varying wireless network with adaptive transmission rates has been provided in~\cite{16}. In~\cite{17}, the problem of allocating power to links as a function of current channel states and queue backlogs to stabilize the system while minimizing the energy expenditure and maintaining low delay in a multiuser network is studied. In~\cite{18}, the problem of designing opportunistic scheduling policies that minimize the average delay in a wireless network with multiple users sharing a wireless channel is studied. In~\cite{19}, energy-efficient scheduling with delay constraints in a multiuser wireless network is studied. The problem of delay minimization under power constraints for uplink transmission in a multiuser wireless network is studied in~\cite{20}. The problem of minimizing the transmission power subject to a delay constraint in a multiuser wireless network is studied in~\cite{21}. However, with the exception of the recent works~\cite{22, 1}, no work has addressed the problem of scheduling in a wireless network with the objective of minimizing the energy consumption and delay using the theory of Whittle index~\cite{23}. In the model in~\cite{22}, at most one user can successfully transmit at a time on the channel. In this paper, we study a wireless network that employs spatial reuse of spectrum, allowing multiple simultaneous transmissions. In~\cite{1}, two Whittle index based stationary policies are provided-- ``Clique Whittle Policy'' and ``Graphical Whittle Policy'' for the same problem. Among the four policies we introduced in this paper, two are non stationary policies. Non stationarity is of interest as using past information for making new decisions may have an advantage of better performance. But as we infer from the simulation results, non stationary policies do not give significantly better performance. The performance of the current four policies is also compared with the policies developed in~\cite{1}. They outperform the policies in~\cite{1} along with some well-known policies in most of the cases.

The  paper is organized as follows. In Section~\ref{section:2}, we describe the model and problem formulation and briefly review the theory of Whittle index. We present four scheduling algorithms based on Whittle-like indices (referred to as `Whittle' indices henceforth) for this problem in Section~\ref{section:3}. We present simulation results in Section~\ref{section:4} and conclude in Section~\ref{section:5}.

\section{Problem formulation and background}\label{section:2}

\subsection{Model and Problem Formulation}\label{section:pf}

\begin{figure}[t]
    \centering
    \includegraphics[height=0.6\linewidth]{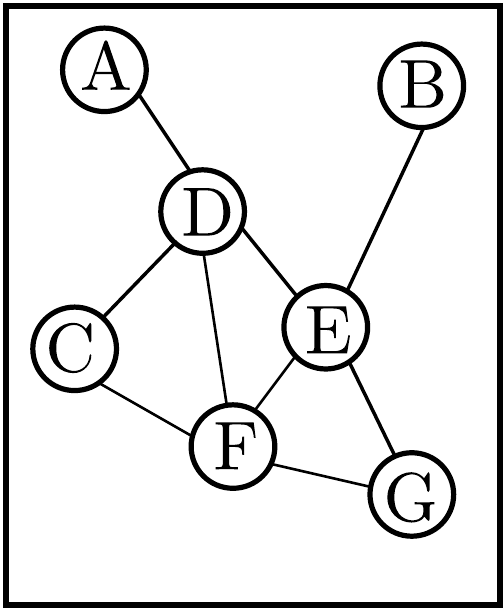}
    \caption{Undirected graph representing a wireless network}
    \label{Net}
\end{figure}
We consider a wireless network consisting of $L$ users deployed in a region and communicating using a single channel. Each user is a transmitter-receiver pair, with a queue at the transmitter of packets to be sent to the receiver. Recall that the wireless medium has the property that simultaneous transmissions by two users that are close to each other interfere with each other, whereas the channel can be simultaneously used at mutually far apart locations without interference. To model this, we represent the network using an undirected graph (fig-\ref{Net}) $\mathcal{G} = (\mathcal{V}, \mathcal{E})$, in which $\mathcal{V}$ is the set of users and there is an edge between two users $i, j \in \mathcal{V}$ iff the transmissions of users $i$ and $j$ interfere with each other.
Let $\mathcal{N}(i)$ be the set of neighbors of user $i$, i.e., the set $\{j \in \mathcal{V} : \exists (i, j) \in \mathcal{E}\}$.

Time is divided into slots of equal durations. The queue of user i evolves according to the dynamics:
\begin{equation}\label{eq-1}
    X_{n+1}^i=[X_n^i-\nu_n^i(X_n^i\land \Psi^i)+\xi_{n+1}^i]\land M^i,
\end{equation}
where $X_n^i$ is the length of the queue of user $i$ in  slot $n$, $\xi_n^i$ is the number of arrivals at the queue of user $i$ in  slot $n$, $M^i$ is the capacity of the buffer of user $i$, $\Psi^i$ is the maximum number of packets that may be transmitted by user $i$ in a slot and $\nu_n^i$ is $1$ if user $i$ transmits in slot $n$ and $0$ otherwise. We say that a user is ``active'' in a slot if it transmits and ``passive'' if not. We assume that the number of packet arrivals, $\xi_n^i$, $n = 0, 1, 2, \dots$, in different slots are independent and identically distributed (IID) random variables with distribution $\mu^i(\cdot)$. 

The cost of holding packets in the queue of user $i$ is $C^i$ per packet per slot. 
The higher the  $C^i$, the more stringent the delay requirements of the packets stored in queue $i$, e.g.,  $C^i$ may be set to a low (resp., high) value if queue $i$ stores elastic traffic such as file transfer packets (resp., real-time traffic such as audio and video flow packets). Let $f^i(z)$ be the ``energy cost'' incurred by user $i$ when it transmits $z$ packets.

Let $\mathcal{N}^*(i) := \mathcal{N}(i)\cup \{i\}$. If two or more users from the set $\mathcal{N}^*(i)$ transmit in time slot $n$, their transmissions interfere with each other, leading to the constraints:
\begin{equation}\label{eq-2}
    \sum_{j\in \mathcal{N}^*(i)}\nu_n^j\leq 1, \forall i.
\end{equation}
If a subset of the users in $\mathcal{V}$ transmits in a time slot subject to (\ref{eq-2}), then that subset constitutes an independent set of nodes in the graph $\mathcal{G} = (\mathcal{V}, \mathcal{E})$. Let $Z^i_n
:= X^i_n \land \Psi^i$. We seek to minimize the time-averaged cost incurred by all users:
\begin{equation}\label{eq-3}
    \lim_{N\uparrow \infty}\frac{1}{N}\sum_{n=0}^{N-1}\sum_{i\in\mathcal{V}}E[\nu_n^if^i(Z_n^i)+C^iX_n^i]
\end{equation}
subject to the interference constraints (\ref{eq-2}). That is, our objective is to select in each slot an independent set of users to activate subject to (\ref{eq-2}),  so as to minimize (\ref{eq-3}). The constraint (\ref{eq-2}) may prevent two mutually non-interfering users from simultaneously transmitting when they have a common interfering user, as shown in~\cite{1}. Nevertheless, to facilitate the following analysis, we impose the constraint  (\ref{eq-2}). In Section~\ref{section:algo}, we provide an algorithm for activating users in different time slots, which ensures that transmitting users form a maximal independent set.

\subsection{Background}\label{section:bg}
We briefly recall here the basics of Whittle index~\cite{23} for cost minimizing restless bandits. Consider a collection of $N \geq 2$ controlled Markov chains ($\approx$ arms of a bandit) $Y^i_n , n \geq 0, i \in {1, \dots , N}$, taking values in discrete state spaces $S^i$, with two modes of operation, active and passive, and corresponding transition probabilities  $p_{i,1}(t|s), p_{i,0}(t|s)$ and running costs $c_1(s), c_0(s)$ resp., where $s, t \in S^i$. The control process associated with $i$th chain is $u_i(n), n \geq 0$, taking values in $\{0, 1\}$ with the interpretation that value $1$ (resp., $0$) corresponds to active (resp., passive) mode. Thus the transition probability at time $n$ for the $i$th process is $p_{i,u_i(n)}(\cdot|Y^i_n)$. The objective is to minimize the average cost 
$$\limsup_{n\uparrow \infty}\frac{1}{n}E[\sum_{m=0}^{n-1}\sum_ic_{u_i(m)}(Y_m^i)]$$
subject to the per stage constraint 
$$\sum_iu_i(n)\leq M, \forall n$$
for some $1 < M < N$, which couples the problems. This constraint makes the problem provably hard~\cite{39}. The Whittle device is to relax it to the average constraint
$$\limsup_{n\uparrow \infty}\frac{1}{n}E[\sum_{m=0}^{n-1}\sum_iu_i(m)]\leq M$$
and consider the unconstrained problem of minimizing
\begin{equation}\label{eq-4}
    \limsup_{n\uparrow \infty}\frac{1}{n}E[\sum_{m=0}^{n-1}\sum_i(c_{u_i(m)}(Y_m^i)+\lambda u_i(m))],
\end{equation}
where $\lambda$ is the Lagrange multiplier. Given $\lambda$, this decouples into individual control problems of minimizing
\begin{equation}\label{eq-5}
    \limsup_{n\uparrow\infty}\frac{1}{n}E[\sum_{m=0}^{n-1}(c_{u_i(m)}(Y_m^i)+\lambda u_i(m))]
\end{equation}
for each $i$. Whittle uses this to motivate the so called Whittle index as follows. The problem is said to be (Whittle) indexable if the set of passive states (i.e., the states $Y^i_m$ for which $u_i(m) = 0$ is the optimal action) for each individual problem $i$ monotonically decreases from the whole state space to the empty set as the `tax' or `negative subsidy' $\lambda$ decreases from $+\infty$ to $-\infty$. If so, the Whittle index for the $i$th problem is the function $\lambda^i : S^i \rightmapsto R$ such that $\lambda^i(s) :=$ the smallest value of $\lambda$ for which both active and passive modes are equally desirable in state $s$. The index rule is then to order, at each time $m$, the current indices $\lambda^i(Y^i_m), 1 \leq i \leq N,$ in decreasing order and render active the processes corresponding to the $M$ lowest indices, breaking ties as per some pre-specified rule, and render passive the remaining $N-M$ processes. One way to motivate this is as follows. The relaxation of per stage constraint to time-averaged constraint means that the actual optimal policy for the latter, i.e., optimal policy for the unconstrained problem with $\lambda =$ the correct Lagrange multiplier, will occasionally cause violation of the per stage constraint. The index rule then goes for an intuitively appealing approximation to it that satisfies the constraint at each time.

\section{Scheduling based on Whittle indices}\label{section:3}

\subsection{Definition of Whittle Indices for our Problem}\label{section:def}

 We relax the constraints (\ref{eq-2}) to the following:
\begin{equation}\label{eq-6}
    \limsup_{n\to \infty}\frac{1}{n}E\left[\sum_{m=0}^n\sum_{j\in \mathcal{N}^*(i)}\nu_n^j\right]\leq 1, \forall i.
\end{equation}
We now use a procedure similar to Whittle's~\cite{23}. In this case, (\ref{eq-4}) gets replaced by
\begin{equation}\label{eq-7}
    \limsup_{n\uparrow \infty}\frac{1}{n}E\left[\sum_{m=0}^{n-1}\sum_i(\nu_m^if^i(Z_m^i)+C^iX_m^i+\lambda^i\sum_{j\in \mathcal{N}^*(i)}\nu_m^j)\right]    
\end{equation}
leading to the individual problems
\begin{equation}\label{eq-8}
    \limsup_{n\uparrow \infty}\frac{1}{n}E\left[\sum_{m=0}^{n-1}\nu_m^if^i(Z_m^i)+C^iX_m^i+\Lambda^i\nu_m^i\right]
\end{equation}
for each $i$, with $\Lambda^i := \sum_{j\in\mathcal{N}^*(i)}\lambda^j$. Treating $\Lambda^i$'s as a surrogate for Whittle tax that is `given', the problems decouple into individual problems and one can employ Whittle’s logic to define a Whittle-like index, for a given state $j$, as that value of $\Lambda^i$ for which the active and passive modes are equally desirable at state $j$. Attractive as this scheme may appear, it is not without problems. There can be a non-trivial loss of information in the sense that the map from ${\lambda^i}$ to ${\Lambda^i}$ may not be invertible. Consider, e.g., a graph with two nodes, say $1$ and $2$, connected by an edge. Then $\mathcal{N}^*(1) = \mathcal{N}^*(2) = \{1,2\}$. So $\Lambda^1 = \Lambda^2 = \lambda^1 + \lambda^2$. Hence in this example, the map from ${\lambda^i}$ to ${\Lambda^i}$ is not invertible.
In concrete terms,  moving over to ${\Lambda^i}$ may effectively change the constraint set itself (see~\cite{1}). 

\subsection{Whittle Index Based Algorithm for Activating Users}\label{section:algo}

In this section, we provide an algorithm for selecting an independent set of users to activate in a given time slot, assuming that the  indices of all the users in the slot have been already computed. In Section~\ref{section:WI comp}, we provide four different approaches for computing the whittle indices-— two non stationary and two stationary methods.

Suppose the indices, $\lambda^i(X^i_n)$ of all the users $i \in V$ have been computed in a given slot $n$. An independent set of users to activate in the slot is selected as follows. First, all users with empty queues are declared passive. Then users $i \in V$ for which $\lambda^i(X^i_n) \leq \lambda^j(X^j_n) \ \forall j \in \mathcal{N}(i)$ are declared active (ties are broken according to some tie-breaking rule, e.g., the user with smaller identifier (ID) is declared active). Next, for every active user $i$, all users $j \in \mathcal{N}(i)$ are declared passive. In the next step, all users $i \in V$  that are not yet declared passive or active for which $\lambda^i(X^i_n) \leq \lambda^j(X^j_n) \ \forall j \in \mathcal{N}(i)$  that are not already declared passive, 
are declared active, and their neighbors are declared passive if already not so. This process is repeated till all users have been declared either active or passive. The set of users that are declared active constitute an independent set. These users transmit in the slot. Furthermore, for implementing the procedure, at any point in time, a user only requires information that is available with its neighboring users and therefore the procedure can be implemented in a distributed manner. Note that this procedure does not satisfy the constraint (\ref{eq-2}), but activates an independent set of users. 

\subsection{Computation of Whittle Index}\label{section:WI comp}
We present four different approaches for computing Whittle indices. Recall the dynamic programming equation for an individual queue $i$~\cite{41}:
\begin{eqnarray}
        \lefteqn{V^i(x^i) = C^ix^i+\min_{\nu_i\in[0,1]}[\nu^if(x^i\wedge \Psi^i) +(1-\nu^i)\Lambda^i} \nonumber \\
        &+\sum_{k}V^i([x^i-\nu^i(x^i\wedge \Psi^i)+k]\wedge M^i)\mu^i(k)] -\beta^i, \label{active}
\end{eqnarray}
where $V^i(\cdot)$ is the value function, $\beta^i$ is the optimal  cost and $\Lambda^i(X^i_n) = \sum_{j\in\mathcal{N}^*(i)}\lambda^j(X^i_n)$. Index $\lambda^i$ for state $X^i_n = x^i$ is calculated in an iterative fashion. The following four methods solve a common linear system of equations in variables $V^i(\cdot),\ \beta^i$, at each iteration, and then update $\lambda^i$s according to an update rule that distinguishes the methods.

\subsubsection{Non Stationary Type-1}\label{section:nos1}
Here  Whittle indices are re-computed in each time slot based on Whittle indices of the preceding slot in  two steps:
\begin{itemize}
    \item Solve the following linear system for  $(V^i(\cdot), \beta^i)$ for every arm $i$:
        \begin{flalign}\label{eq-10}
            \begin{split}
                V^i(y^i) = &\sum_{k}V^i([y^i-y^i\wedge \Psi^i+k]\wedge M^i)\mu^i(k)-\beta^i\\
                &+C^iy^i+f(y^i\wedge \Psi^i),\ y^i\geq x^i,
            \end{split}\nonumber\\[7pt]
            \begin{split}
                V^i(y^i) =&\sum_{k}V^i([y^i+k]\wedge M^i)\mu^i(k)+\sum_{j\in \mathcal{N}^*(i)}\lambda_{t-1}^j\\
                &-\beta^i+C^iy^i,\ x^i>y^i \neq 0, 
            \end{split}\nonumber\\[7pt]
            V^i(0) = 0.
        \end{flalign}
    \item Compute the new Whittle indices using the following $\lambda$ iteration,
        \begin{eqnarray}
                \lefteqn{\lambda^{i}_{t} = \lambda^{i}_{t-1} + \gamma[f^i(x^i\wedge \Psi^i)-\boxed{\lambda^i_{t-1}}} \nonumber\\
                &&+\sum_{k}\mu^i(k)(V^i([x^i-x^i\wedge \Psi^i+k]\wedge M^i)] \nonumber \\
&&-V^i([x^i+k]\wedge M^i). \label{eq-11} 
            \end{eqnarray}
\end{itemize}
    
\subsubsection{Non Stationary Type-2}\label{section:nos2}
 As above, but replace (\ref{eq-11}) by
        \begin{flalign}\label{eq-13}
            \begin{split}
                \lambda^{i}_t = &\lambda^{i}_{t-1} + \gamma[f^i(x^i\wedge \Psi^i)-\boxed{\sum_{j\in \mathcal{N}^*(i)}\lambda^j_{t-1}}\\
                &+\sum_{k}\mu^i(k)(V^i([x^i-x^i\wedge \Psi^i+k]\wedge M^i)\\
                &-V^i([x^i+k])\wedge M^i)].
            \end{split}
        \end{flalign}

\subsubsection{New Stationary Type-1}\label{section:nes1}
This policy is a stationary policy, so the Whittle indices can be computed ahead of system/ simulation start. Here we perform a few iterations to compute  the $\lambda$s  to estimate the actual Whittle indices as follows.  Initialize all $\lambda$s to zero. Then 
at each step, solve the above linear system  and update the indices according to
        \begin{flalign}\label{eq-15}
            \begin{split}
                \lambda^{i}_n = &\lambda^{i}_{n-1} + \gamma[f^i(x^i\wedge \Psi^i)-\boxed{\lambda^i_{n-1}}\\
                &+\sum_{k}\mu^i(k)(V^i([x^i-x^i\wedge \Psi^i+k]\wedge M^i)\\
                &-V^i([x^i+k])\wedge M^i)].
            \end{split}
        \end{flalign}

After executing these steps a number of times, the final  $\lambda$s are taken to be the indices for the given state of the system.

\subsubsection{New Stationary Type-2}\label{section:nes2}
As above except that the $\lambda$ iteration is replaced by
        \begin{flalign}\label{eq-17}
            \begin{split}
                \lambda^{i}_n = &\lambda^{i}_{n-1} + \gamma[f^i(x^i\wedge \Psi^i)-\boxed{\sum_{j\in \mathcal{N}^*(i)}\lambda^j_{t-1}}\\
                &+\sum_{k}\mu^i(k)(V^i([x^i-x^i\wedge \Psi^i+k]\wedge M^i)\\
                &-V^i([x^i+k])\wedge M^i)].
            \end{split}
        \end{flalign}

\subsubsection{Explanation for Above Computational Schemes}
The linear system (\ref{eq-10}) above constitutes the Poisson equation (i.e., constant policy dynamic programming equation) for the chain controlled by a fixed stationary Markov policy, viz., the threshold policy with threshold $x^i$. Under irreducibility hypothesis (more generally, uni-chain property, see, e.g.,~\cite{41}), this has a solution $(V^i, \beta^i)$ where $\beta^i$ is uniquely given as the average cost under this policy and the `value function' $V^i$ is unique up to an additive constant. The additional condition $V^i(0) = 0$ then renders it unique.
The iterates in the four approaches (\ref{eq-11}), (\ref{eq-13}), (\ref{eq-15}), (\ref{eq-17}) are also similar except for the terms present in a box. They make incremental adjustments in $\lambda^i$s so as to force the defining equality for the Whittle index.

\section{Simulations}\label{section:4}
In this section, we evaluate the performances of the proposed algorithms and compare them with those of the policies proposed in~\cite{1}, the well known Slotted ALOHA~\cite{kurose2005computer}, maximum-weight scheduling (MWS)~\cite{182479}, quadratic Lyapunov drift~\cite{1650347} algorithms, and an algorithm recently proposed by Cella and Cesa-Bianchi~\cite{38}, via simulations (see our code \cite{code} for details). The performance is evaluated in terms of two metrics-- average cost and average total number of packets dropped per time slot-– at all the users in the network.  Note that penalizing queue length automatically penalizes packet drops. Putting an explicit penalty on packet drops adds an extra cost for the state (the buffer length),  a cost that retains the monotonicity and (discrete) convexity properties of the running cost and does not affect our heuristic argument leading to the proposed policy. It is expected to decrease the mean packet drops at the expense of mean energy cost. This is a topic for future research.

\subsection{Simulation Model}
In our simulations \cite{code}, we consider the model  in Section~\ref{section:pf} with $L = 20$ users and buffer capacity $M^i = 100$ for
each user $i \in V$. The location of each user is selected uniformly at random in a square of dimensions $1\times 1$ unit. Two users are neighbors iff the distance between them is less than a threshold $d$, which is a parameter. Unless otherwise mentioned, we use $d = 0.6$ units throughout the simulations. Let $\Psi^i$ be the maximum number of packets that may be transmitted by user $i$ in a given slot. For the index based algorithms, we consider two cases: (i) $\Psi^i = \infty$, and (ii) $\Psi^i$ is uniformly distributed between $1$ and $M^i/5$ for user $i$, independent of other users. We refer to cases (i) and (ii) as the ``unrestricted transmission'' and ``restricted transmission'', respectively. Note that in case (i), a user that transmits in a time slot sends all the packets in its queue. In our simulations, unless otherwise mentioned, under the Slotted ALOHA, Max-Weight Scheduling, quadratic Lyapunov drift and Cella and Cesa-Bianchi's algorithms, the value $\Psi^i$ for user $i$ is the same as that in the restricted case of the index based algorithms. We assume that the number of packets $\xi^i_n$ that arrive at user $i$ in time slot $n$ is a Poisson random variable with mean $l^i$. Also, unless otherwise mentioned, $l^i$ is selected uniformly at random to be between $1$ and $M^i/10$ for each $i$, independent of other users. We use $\theta = 200$ in the quadratic Lyapunov drift algorithm.

\subsection{Simulation Results}
\begin{figure}[H]
    \begin{subfigure}{0.45\textwidth}
        \centering
        \includegraphics[width=\textwidth,height=0.6\textwidth]{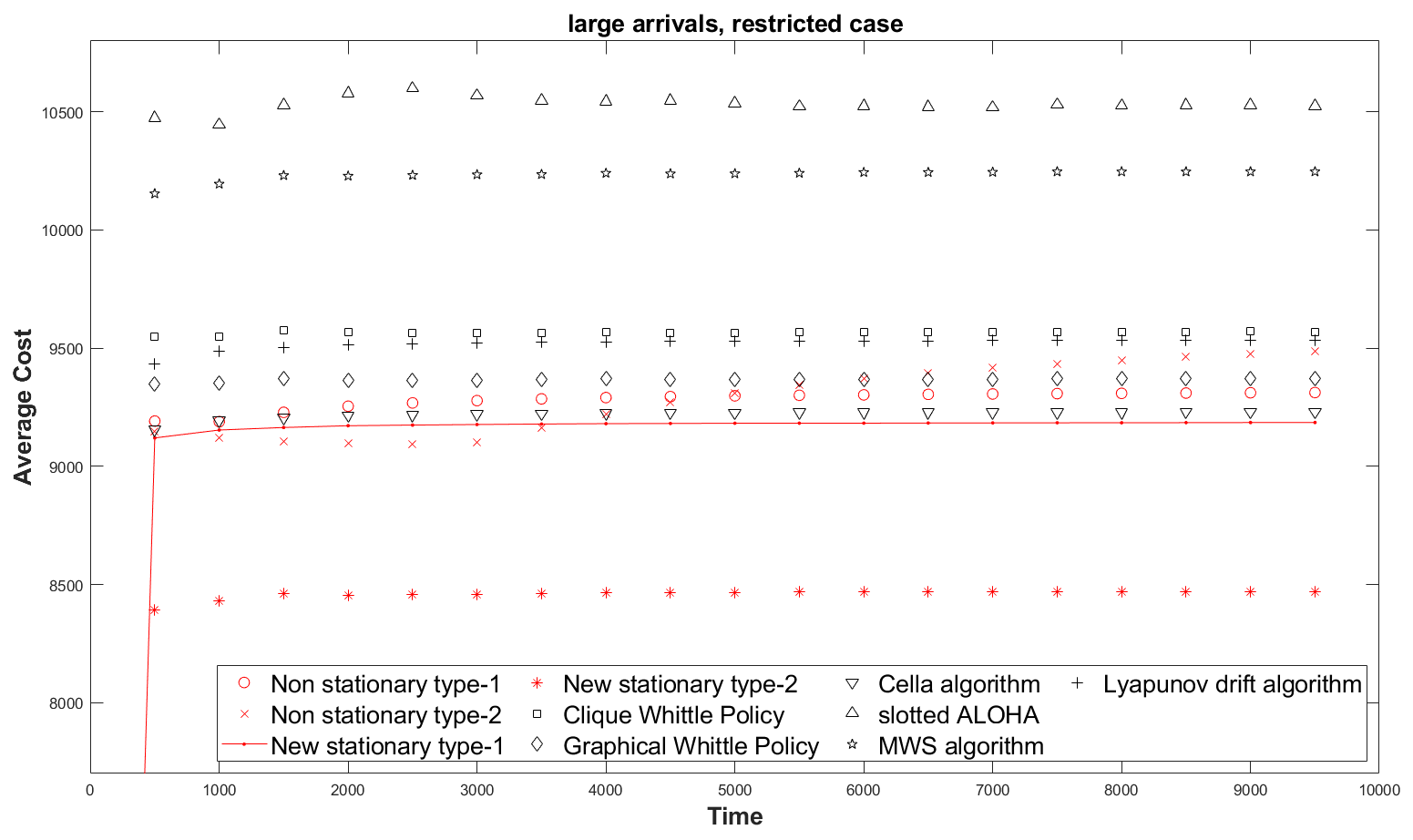}
        \caption{Large arrival rates, restricted transmissions}
        \label{fig:AC-lr}
    \end{subfigure}\hfill
    \begin{subfigure}{0.45\textwidth}
       \centering
        \includegraphics[width=\textwidth,height=0.6\textwidth]{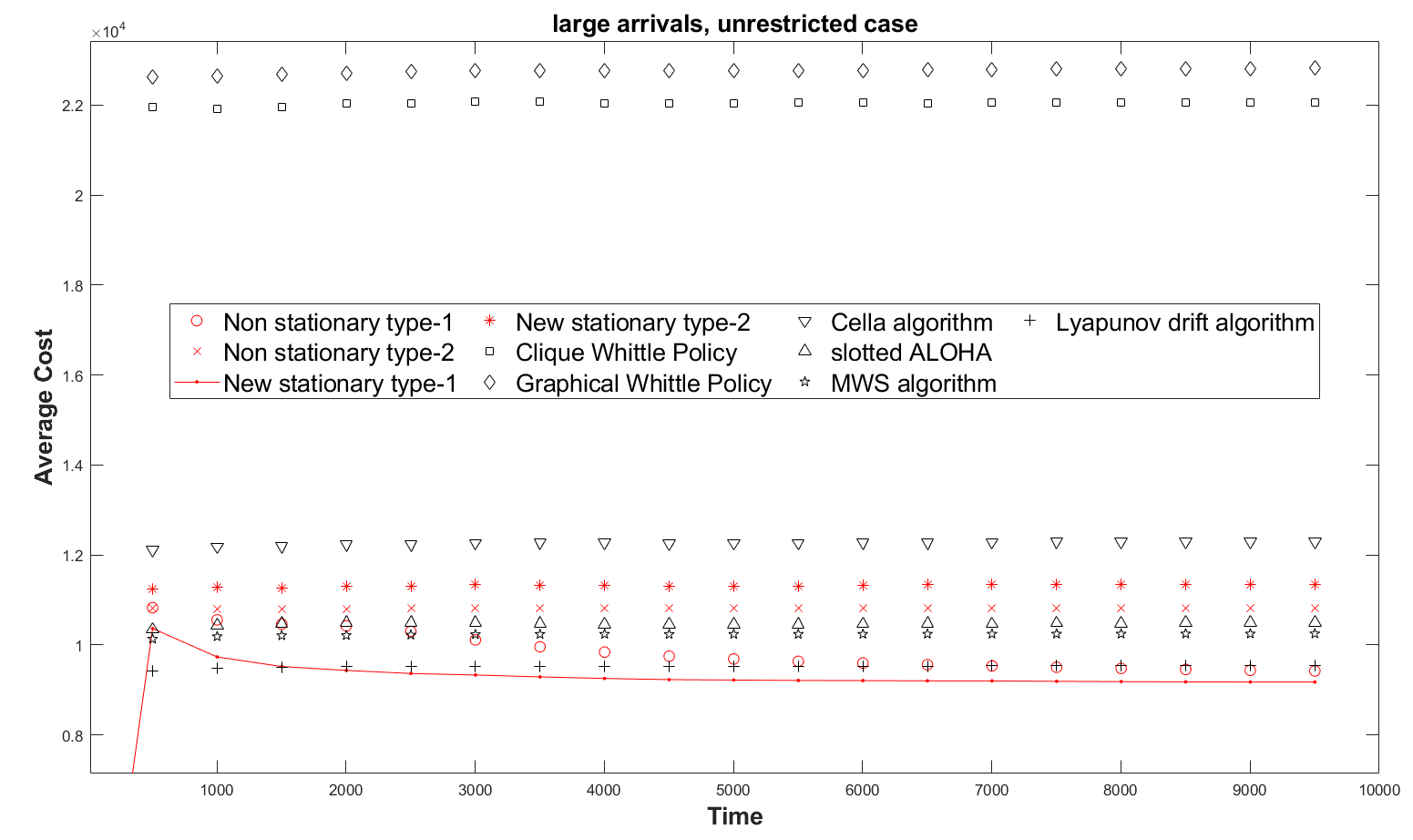}
        \caption{Large arrival rates, unrestricted transmissions}
        \label{fig:AC-lur}
    \end{subfigure}\hfill
    \begin{subfigure}{0.45\textwidth}
           \centering
        \includegraphics[width=\textwidth,height=0.6\textwidth]{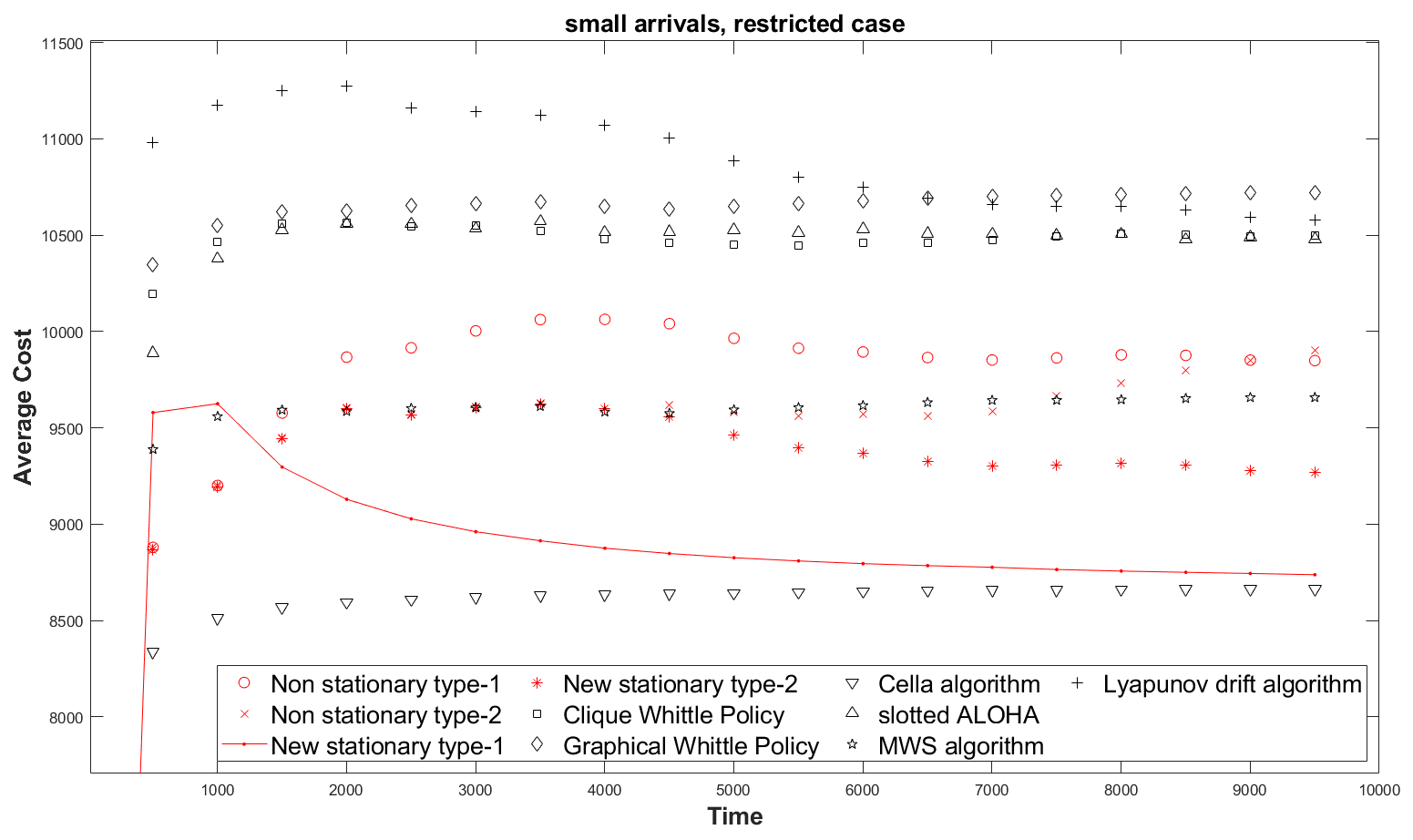}
        \caption{Small arrival rates, restricted transmissions}
        \label{fig:AC-sr}
    \end{subfigure}\hfill
    \begin{subfigure}{0.45\textwidth}
       \centering
        \includegraphics[width=\textwidth,height=0.6\textwidth]{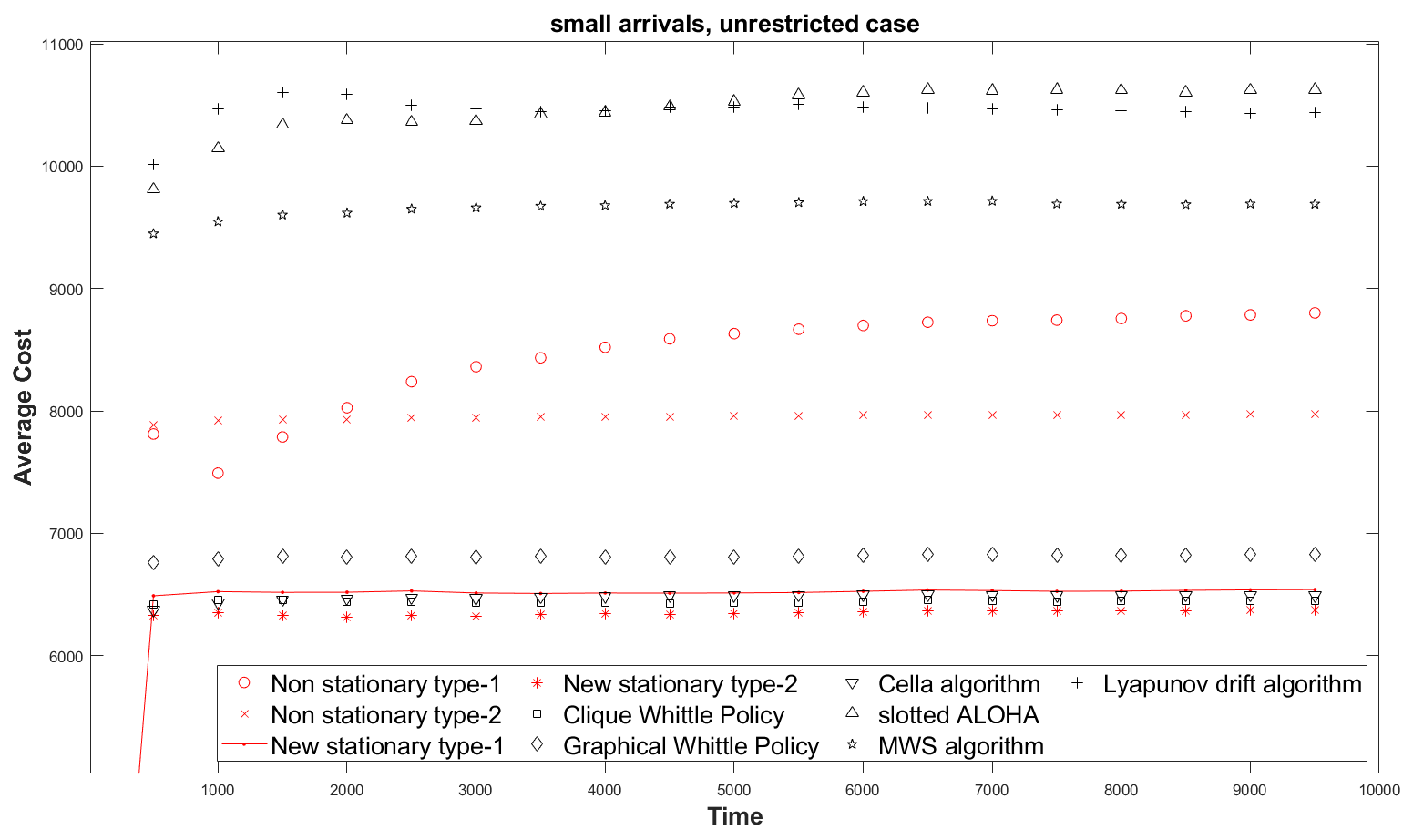}
        \caption{Small arrival rates, unrestricted transmissions}
        \label{fig:AC-sur}
    \end{subfigure}
    \caption{Average Cost Comparison}
\end{figure}

\begin{figure}[H]
    \begin{subfigure}{.45\textwidth}
        \centering
        \includegraphics[width=\textwidth,height=0.6\textwidth]{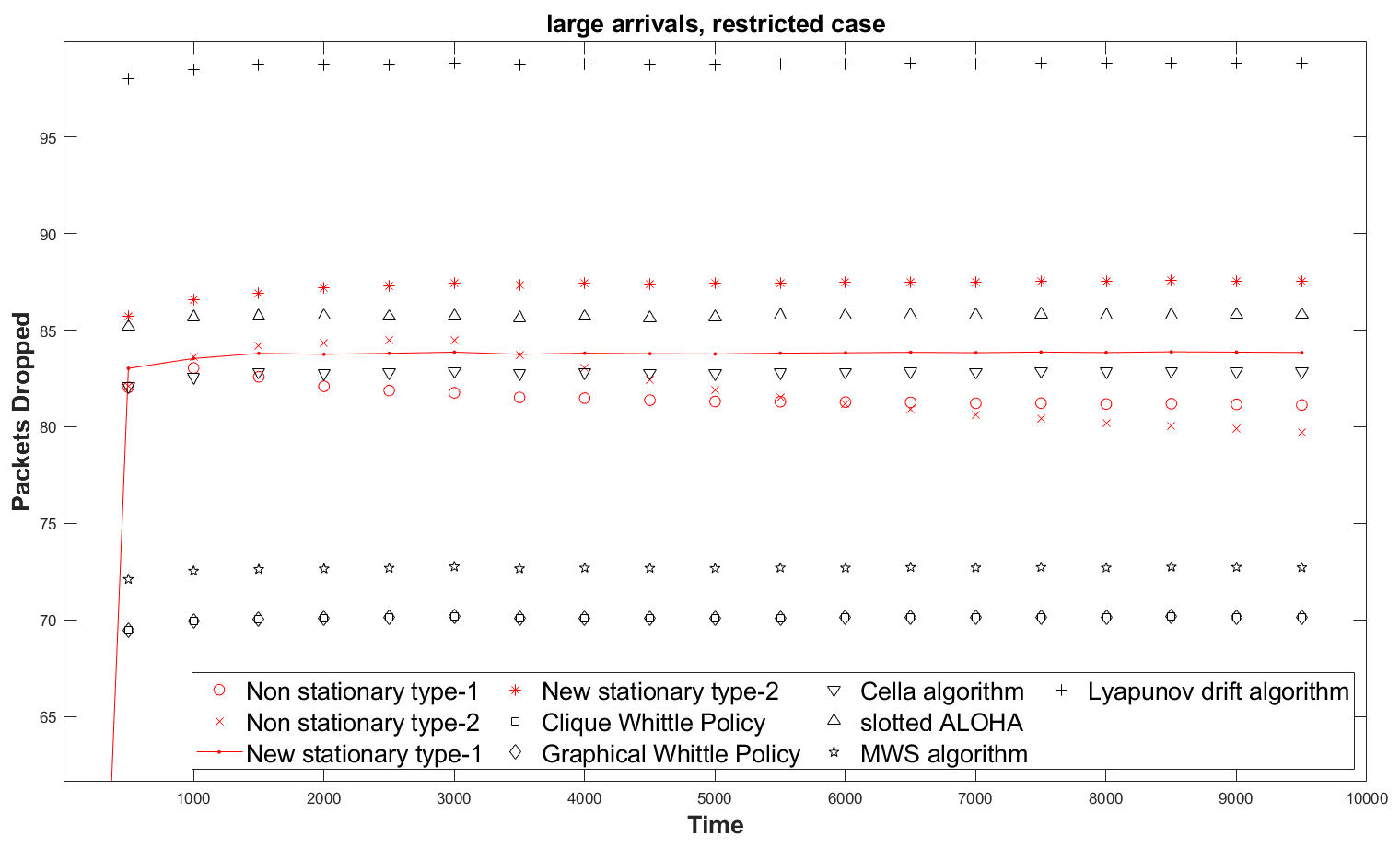}
        \caption{Large arrival rates, restricted transmissions}
        \label{fig:PD-lr}
    \end{subfigure}\hfill
    \begin{subfigure}{.45\textwidth}
        \centering
        \includegraphics[width=\textwidth,height=0.6\textwidth]{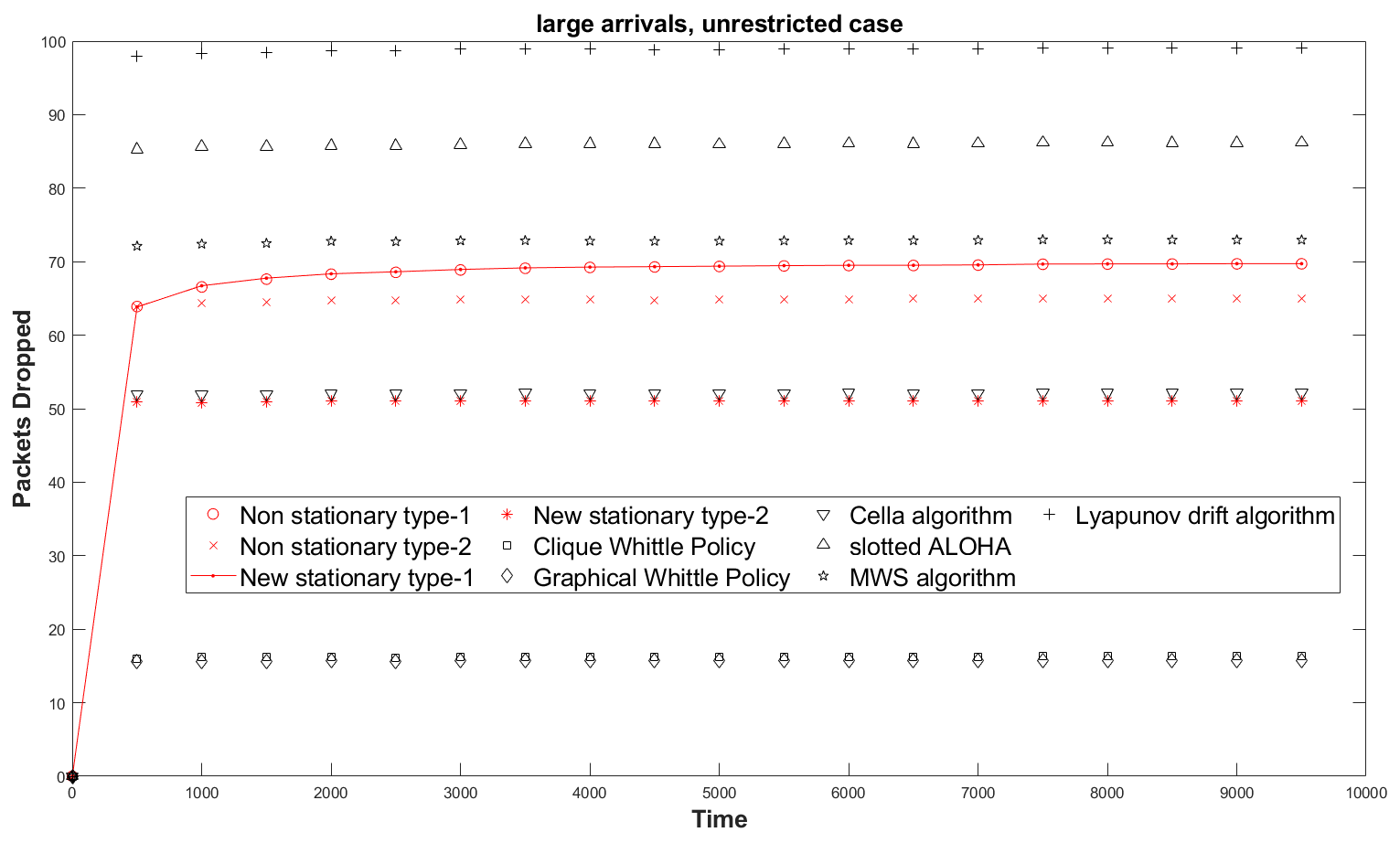}
        \caption{Large arrival rates, unrestricted transmissions}
        \label{fig:PD-lur}
    \end{subfigure}\hfill
    \begin{subfigure}{.45\textwidth}
        \centering
        \includegraphics[width=\textwidth,height=0.6\textwidth]{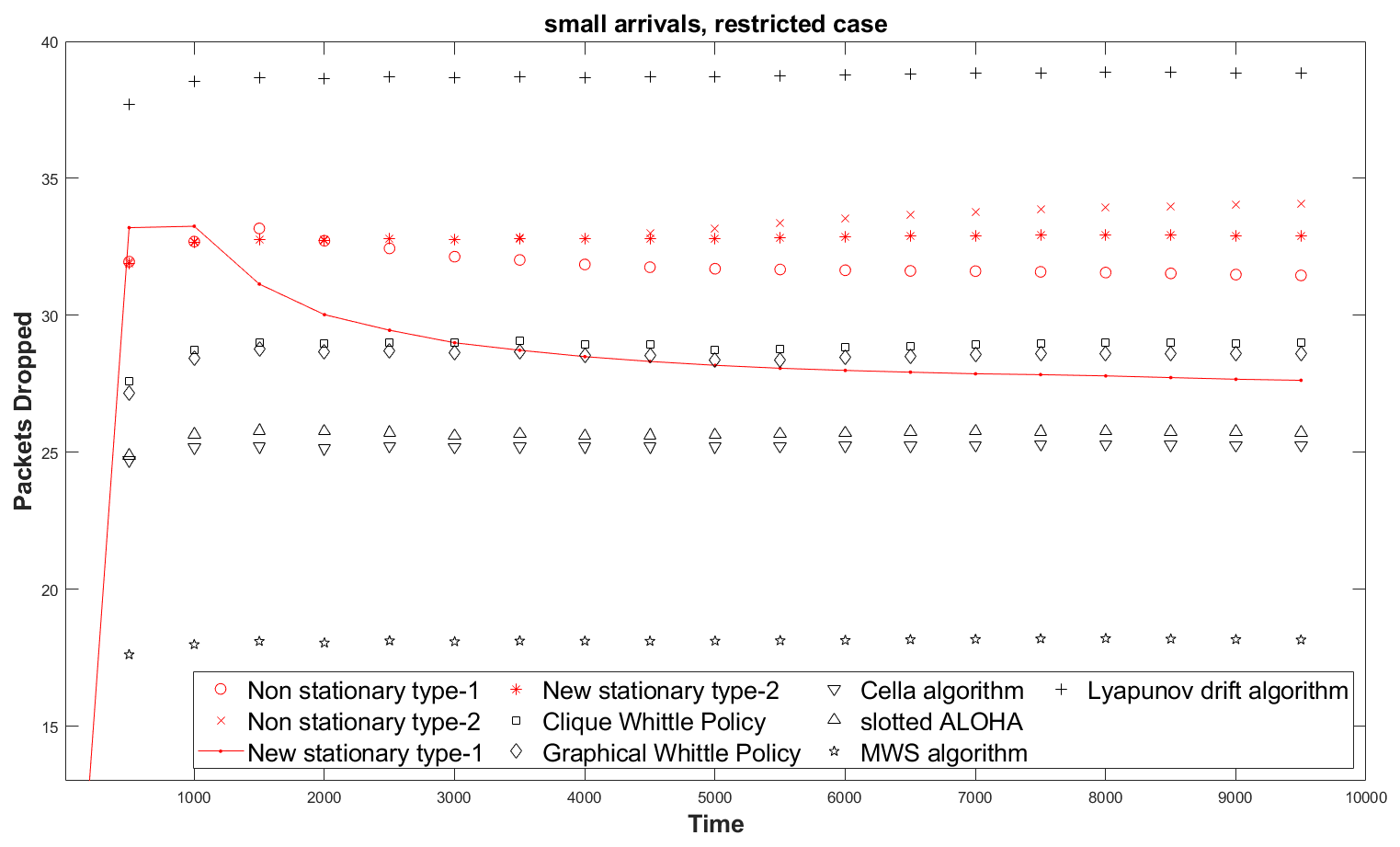}
        \caption{Small arrival rates, restricted transmissions}
        \label{fig:PD-sr}
    \end{subfigure}\hfill
    \begin{subfigure}{.45\textwidth}
       \centering
        \includegraphics[width=\textwidth,height=0.6\textwidth]{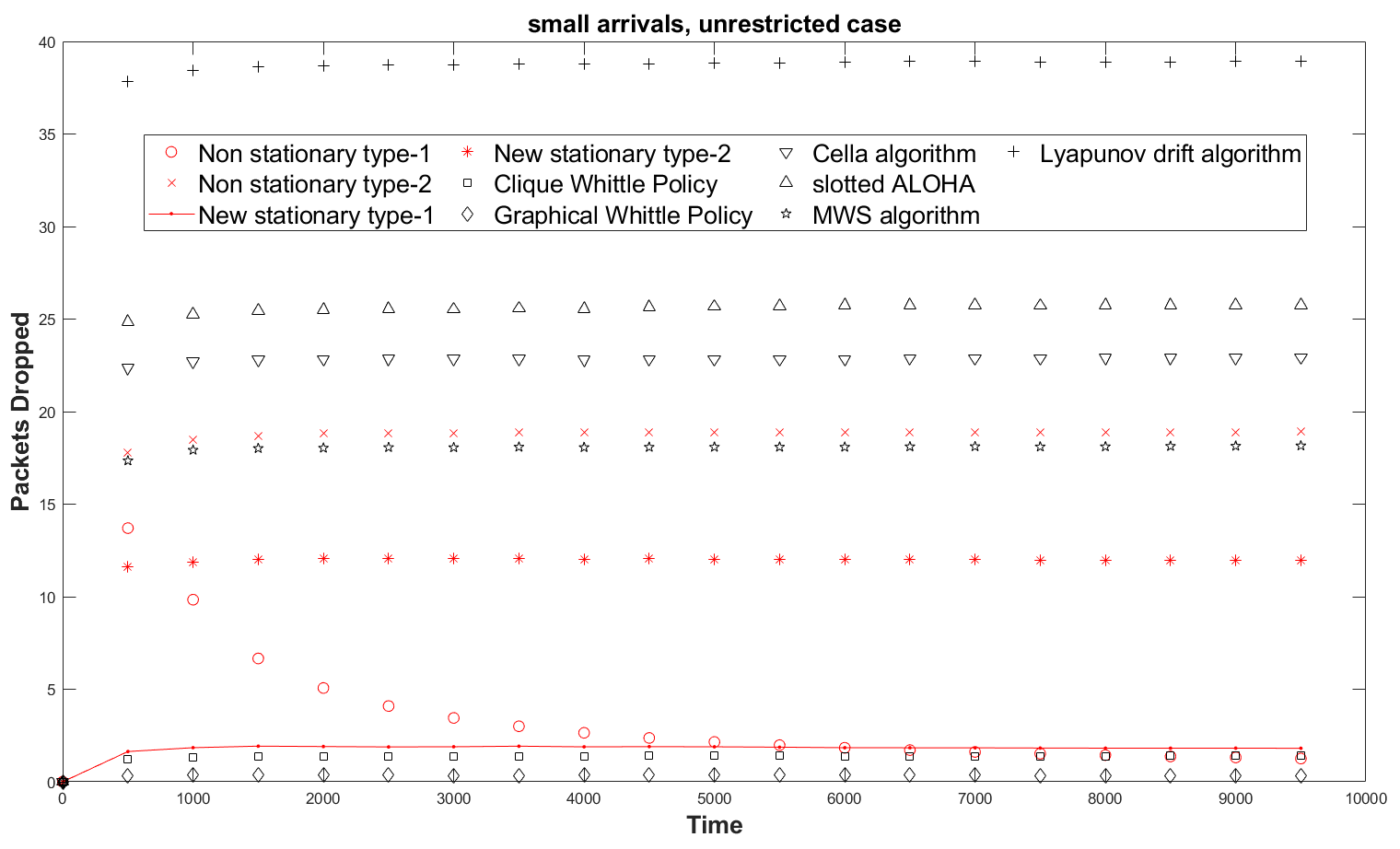}
        \caption{Small arrival rates, unrestricted transmissions}
        \label{fig:PD-sur}
    \end{subfigure}
    \caption{Average Packets Dropped Comparison}
\end{figure}

In the plots below, by ``large arrival rates'' (respectively, ``small arrival rates''), we mean that $l^i$ is chosen uniformly at random between $1$ and $M^i/8$ (respectively, between $1$ and $M^i/15$). 
Figs.~\ref{fig:AC-lr},~\ref{fig:AC-lur},~\ref{fig:AC-sr} and~\ref{fig:AC-sur} (respectively,~\ref{fig:PD-lr},~\ref{fig:PD-lur},~\ref{fig:PD-sr} and~\ref{fig:PD-sur}) compare the performances of Non stationary type-1 policy, Non stationary type-2 policy, New stationary type-1 policy, New stationary type-2 policy, the  policies in~\cite{1}-- ``Clique Whittle Policy'' and ``Graphical Whittle Policy'', Slotted ALOHA, Max Weight Scheduling (MWS), quadratic Lyapunov drift algorithm, and Cella and Cesa-Bianchi's algorithm in terms of average cost (respectively, average total number of packets dropped by all  users in the network per time slot), for the cases with  large/ small arrival rates and restricted/ unrestricted transmissions, respectively. 

As per the average cost  (Figs.~\ref{fig:AC-lr},~\ref{fig:AC-lur},~\ref{fig:AC-sr} and~\ref{fig:AC-sur}), the New stationary policies: type-1 and type-2 outperformed all the other policies except for the case of small arrival rates with restricted transmissions where the Cella and Cesa-Bianchi's algorithm is the optimal, but there is just a marginal difference between the New stationary policies and the latter. As per the packets dropped (Figs.~\ref{fig:PD-lr},~\ref{fig:PD-lur},~\ref{fig:PD-sr} and~\ref{fig:PD-sur}), the proposed stationary policies do not have an upper hand and there isn't one best policy, i.e., optimal in all cases. The Whittle policies in~\cite{1} relatively have better performance by comparison in all cases. 
We consider the preferred policy to be the New stationary policy type-1 as it showed better performance in many cases.  The intuition behind the observed results is that resetting the $\lambda$ iterates to zero as in the new stationary policies removes the undue influence of the past. Similarly, having the term in the box depend on the neighboring $\lambda$'s introduces a `competitive' aspect in the coupling between neighboring nodes and worsens the performance. A detailed analysis of these phenomena are a subject for future work.



\section{Conclusions and future work}\label{section:5}
We proposed four Whittle index based scheduling policies for scheduling packet transmissions with the objective of minimizing the energy consumption and data transmission delay of users in a wireless network in which spatial reuse of spectrum is employed. Two of the policies are non stationary and  two are stationary. We evaluated the performance of these policies via extensive simulations and showed that they outperform the well-known Slotted ALOHA, maximum-weight scheduling, quadratic Lyapunov drift, algorithm proposed by Cella and Cesa-Bianchi and the two Whittle index based policies (Clique and Graphical Whittle policies) in~\cite{1}. A potential direction for future research is to develop analytical bases for these schemes and their improved variants. Another direction for future work is to extend the results of this paper to the case where the users of the network use millimeter wave (mmWave) spectrum and each transmitter-receiver pair uses directional transmissions (beamforming) for exchanging information \cite{singh2022user}. \\

\noindent \textbf{Acknowledgments} VSB was supported by a S.\ S.\ Bhatnagar Fellowship from the Govt.\ of India. The work of
GSK was supported in part by the project with code RD/0121-MEITY01-001.

\printbibliography[heading=bibintoc]

\newpage

\end{document}